# $H_2 + H_2O \rightarrow H_4O$: Synthesizing Hyper-hydrogenated Water in Small-sized Fullerenes?

*Endong Wang,[1,2] Yi Gao [1*]*

[1] School of Chemistry and Chemical Engineering, Liaoning Normal University, Dalian 116029, China

[2] Shanghai Advanced Research Institute, Chinese Academy of Sciences, Shanghai 201210, China

AUTHOR INFORMATION

**Corresponding Author**

*Authors to whom correspondence should be addressed: gaoyi@zjlab.org.cn

ABSTRACT

Nanoscale confinement provides an ideal platform to rouse some exceptional reactions which cannot happen at the open space. Intuitively, $H_2$ and $H_2O$ cannot react. Herein, through utilizing small-sized fullerenes ($C_{24}$, $C_{26}$, $C_{28}$, and $C_{30}$) as nanoreactors, we demonstrate a hyper-hydrogenated water species, $H_4O$, can be easily formed using $H_2$ and $H_2O$ at ambient condition by *ab initio* molecular dynamics simulations. The $H_4O$ molecule rotates freely in the cavity of the cages and maintains its structure during the simulations. Further theoretical analysis indicates $H_4O$ in the fullerene possesses the high stability thermodynamically and chemically, which can be rationalized by the electron transfer between $H_4O$ and the fullerene. This work highlights the possibility of fullerene as nanoreactor to provide confinement constraint for unexpected chemistry.

**TOC GRAPHICS**

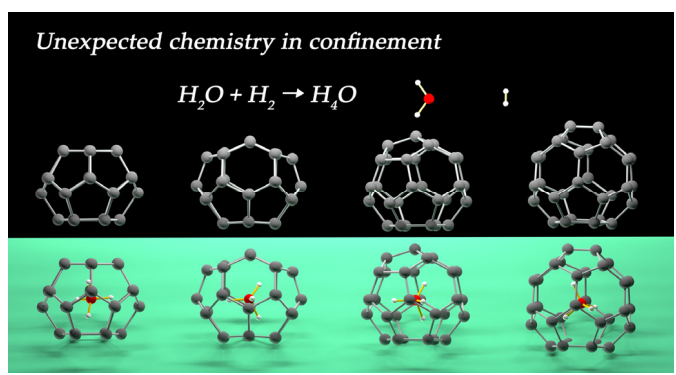

1. INTRODUCTION

Nanoscale confinement has been extensively recognized as a special environment for the distinct physicochemical phenomena, including the increased catalytic activities, enhanced protein folding, and the superfast water transport. For example, Bao et al. found that the catalytic performance of the single atom catalyst can be enhanced when confined in two-dimensional materials.[1] Shrestha et al. reported 100 times faster folding rate of G-quadruplex in DNA origami nanocages compared with the case of diluted or molecularly crowded buffer solutions.[2] Through confining water into charges arranged nanotube, water can be pushed from one end to another like a molecular water pump.[3]

In particular, fullerenes can provide the closed-space confinement cavity, which have been used to encapsulate readily available molecules (such as $H_2$, $H_2O$ et al.) to exhibit the distinct properties.[4-6] Kurotobi and Murata employed $C_{60}$ to accommodate a $H_2O$ molecule with the rapid rotation inside.[7] A later study indicated $F_2^-@C_{60}^+$ owns substantially longer F-F bond than that for free $F_2$ due to the electron-transfer between the core-shell.[8] Murata et al. prepared $(H_2O \cdot HF)@C_{70}$ and found no proton transfer proceed even at 140°C.[9] In addition, more species as $CH_4@C_{60}$,[10, 11] $H_2O@C_{60}$,[10, 12] superalkali@$C_{60}$,[13] $X@C_{59}N$ ($X = H_2O$, $H_2$),[14] $He@C_{60}$ isomer,[15] $Ne@C_{60}$ isomer,[15] and $CHFClBr@C_{82}$ isomer[16] were also obtained in respective confined space. Above significant progress stimulates us to consider the possibility of synthesizing non-existent molecule in the confinement environment.

In this work, we employed the *ab initio* molecular dynamics simulations to show the direct reaction of $H_2O$ with $H_2$ to form $H_4O$ at the ambient condition using small-sized fullerenes ($C_{24}$, $C_{26}$, $C_{28}$, and $C_{30}$) as the nanoreactor. The radii of the fullerenes employed range from 2.3 Å to 2.6 Å, and the O-H bond length is around 0.96 Å, which can ensure the enough space to accommodate $H_2O$. The stability of $H_4O@C_{24}$, $H_4O@C_{26}$, $H_4O@C_{28}$, and $H_4O@C_{30}$ were further verified through frequency calculation, HOMO-LUMO gap, and additional AIMD simulations.

## 2. COMPUTATIONAL METHODS

AIMD simulations reported in this work were performed using the Quickstep module implemented in the CP2K-6.1 package.[17] PBE0 functional coupled with Grimme's D3 dispersion correction was applied in the simulations. Goedecker−Teter−Hutter pseudopotentials[18] and a double-ζ valence plus polarization (DZVP-MOLOPT-GTH) basis set are used. The plane wave cutoff is 300 Ry. In all the AIMD simulations, the time step is 0.5 fs. The cubic simulation cell under periodic boundary condition with a cell length of 15.0 Å was applied. The temperature was controlled by a Nose thermostat with a target temperature of 298 K. The simulation utilizes NVT ensemble and was conducted for 15 *ps*. Unless otherwise specified, data for the last 10 *ps* were collected for further analysis.

The geometrical optimization and frequency calculations of the electronic structure calculations were conducted using Gaussian 09-D01 package.[19] Harmonic vibrational analysis with no imaginary frequency was done to characterize $H_4O@C_{24}$, $H_4O@C_{26}$, $H_4O@C_{28}$, and $H_4O@C_{30}$ as minima. Natural population analysis (NPA) was done via the NBO 3.0 module implemented in Gaussian 09-D01 package.

Localized molecular orbitals (LMOs) were obtained through two separate methods including Foster-Boys localization method and Pipek-Mezey localization method implemented in Orca 5.0.3 package.[20] PBE0-D3 functional, triple-ζ basis set (def2-TZVP),[21] and the auxiliary basis set (def2/J)[22] were applied during orbital localization.

Energy decomposition analysis were performed at PBE0-D3/def2-TZVP level of theory using a canonical molecular orbital energy decomposition analysis (CMO-EDA) approach[23] though GAMESS-US software, version 11 No. 2017.[24]

3. RESULTS AND DISCUSSION

Fig. 1 gives the AIMD trajectory of $H_4O$ formation with the two different initial configurations of $H_2$ and $H_2O$ in $C_{24}$ at 298 K. The first configuration refers to that $H_2$ resides near the H atoms of the $H_2O$ molecule at 0 *fs*. As the reaction begins, Fig. 1(a) shows that the bond length of O-$H_1$ and O-$H_2$ slightly increase until around 10 *fs*. During this period, the H-H bond of the $H_2$ molecule gradually elongates from 0.60 Å to 1.05 Å. At around 10.5 *fs* ($t_2$), H-H bond of $H_2$ breaks and one H approaches to $H_2O$ to form O-$H_3$ with the increased bond length of 1.10 Å. Then, the length of O-$H_1$, O-$H_2$, and O-$H_3$ decrease until $H_3O$ was formed at around 20 *fs*. $H_3O$ is a metastable species lasting ~15 fs, during which we observed the exchange of $H_4$ atom with $H_1$ atom. After that, $H_3O$ gradually evolves to $H_4O$ using around 100 *fs* and keeps in stable until 1 *ps*. Fig. 1(b) gives the trajectory of another initial configuration with $H_2$ locating near the lone electron pair of $H_2O$. Similar as Fig. 1(a), Fig. 1(b) shows that $H_2$ decomposes first. At 9.5 *fs*, the two H atoms of $H_2$ molecule own the largest distance of 2.60 Å, which indicates the full dissociation of the H-H bond. Both H atoms move rather near to the O atom at around this time point. At 24.5 *fs*, it shows that $H_4$ atom was captured by the O atom although O-$H_4$ length continues to fluctuate in the following reaction steps. At 30 *fs*, O-$H_3$ bond tends to be formed. Finally, $H_4O@C_{24}$ was obtained at around 100 *fs*, which is still stable until 1 *ps*. To further show the detailed process of the formation of $H_4O@C_{24}$, the interactions among $C_{24}$, $H_2O$, and $H_2$ extracted from the initial structure of AIMD simulation were evaluated. Table S1 show that the interactions between $C_{24}$ and $H_2$ (17.13 and 14.48 eV) are larger than those between $C_{24}$ and $H_2O$ (5.67 and 8.62 eV) for both AIMD initial structures. Both interactions are larger than the interaction between $H_2O$ and $H_2$ (2.55 and 1.00 eV), which means the confined $C_{24}$ environment pushes $H_2O$ and $H_2$ to each other. The generation of $H_4O$ from $H_2O$ and $H_2$ were verified through direct optimization at the PBE0-D3/def2tzvp level of theory. Figure S1 suggests that this process pairs with the decrement of total energy. The approach of $H_2O$ and $H_2$ accompanies the dissociation of H-H bond, which can be understood from the bond dissociation energy of H-H bond of $H_2$ (4.52 eV calculated at PBE0-D3/def2tzvp) is smaller than O-H bond of $H_2O$ (5.19 eV calculated at PBE0-D3/def2tzvp), thus breaks more easily.

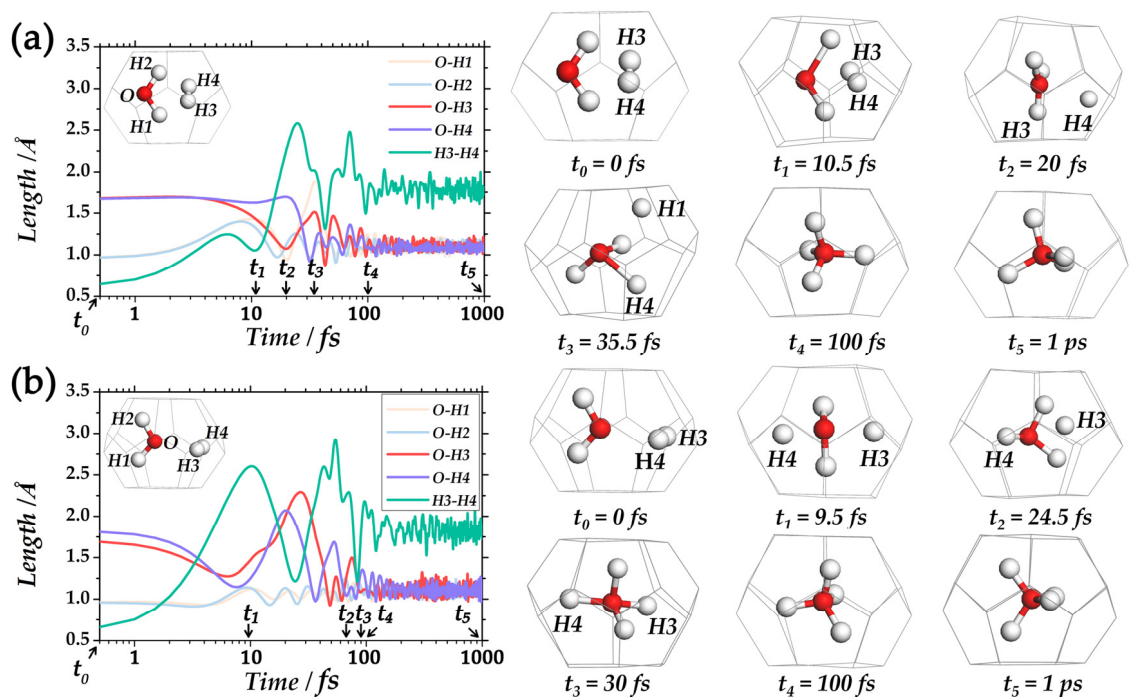

**Fig. 1** Variations of key bond length and important structures for the first initial configuration (a), for the other initial configuration (b) for the generation of $H_4O$ from $H_2O$ and $H_2$. To be clear, logarithmic scale was applied in the x-coordinate.

To verify the thermal stability of $H_4O@C_{24}$, the AIMD simulation at 298 K was further conducted. The simulation was conducted for 15 *ps* and the data of the last 10 *ps* were collected for analysis. Fig. 2 shows that the four O-H bonds including O-$H_1$, O-$H_2$, O-$H_3$, and O-$H_4$ fluctuate around a reasonable range of 1.0-1.25 Å, which demonstrates the structure of $H_4O$ species maintains within $C_{24}$. Notably, $H_4O$ does not hold the still orientation relative to the outer $C_{24}$ cage during the simulation. The change of the length of $H_1$-$C_1$ between around 1.5 Å and 3.2 Å suggests that $H_4O$ species keeps rotation within $C_{24}$, which is similar to the experimental observation of $H_2O$ in $C_{60}$.[7]

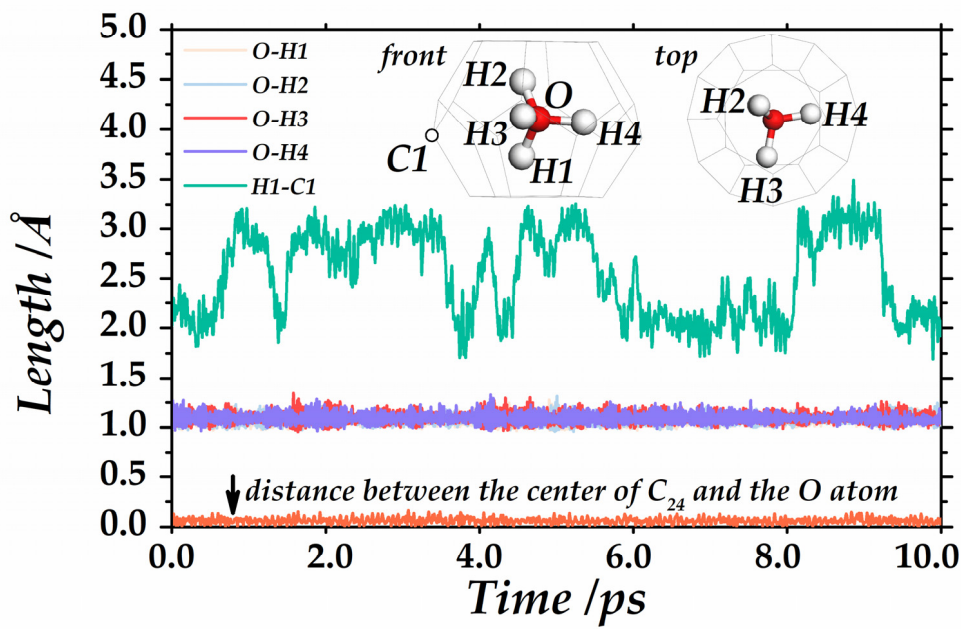

**Fig. 2** Fluctuations for key bond lengths of $H_4O@C_{24}$ during AIMD simulation.

The optimized structure of $H_4O@C_{24}$ via PBE0-D3/def2tzvp is shown in the inset of Fig. 2. Calculations based on PBE0-D3/def2tzvp, PBE0-D3/cc-pvdz, M06-2X-D3/def2tzvp, and wB97XD/def2tzvp were used to verify stability of $H_4O@C_{24}$. As in Fig. 2, the four O-H bonds present simultaneously indicating an over-coordinated status for the center O atom, which is similar to the C atom of $[CTi_7^{2+}]$.[25] This result was confirmed at the levels of PBE0-D3/cc-pVTZ, M06-2X-D3/def2tzvp, and wB97XD/def2tzvp. Harmonic vibrational analysis with no imaginary frequency was done to characterize $H_4O@C_{24}$. The lowest frequencies were 180.97 cm$^{-1}$ for PBE0-D3/def2tzvp, 141.97 cm$^{-1}$ for PBE0-D3/cc-pVTZ, 175.29 cm$^{-1}$ for M06-2X-D3/def2tzvp, and 156.27 cm$^{-1}$ for wB97XD/def2tzvp. All values are positive suggesting the H$_4$O@C$_{24}$ as a local minimum. The HOMO-LUMO gap of H$_4$O@C$_{24}$ is 2.05 eV at the PBE0-D3/def2tzvp level of theory, which indicates the high chemical stability of this species. To further show the structural characteristic of H$_4$O in C$_{24}$, major geometrical parameters calculated via PBE0-D3/def2tzvp are given in Table 1. H$_4$O owns slightly larger (0.996, 0.997, 1.021, 1.021 Å) O-H length when comparing with the O-H length of H$_2$O (0.962 Å). Six H-O-H angles are near to 109.28°, which is caused by the valence shell electron pair repulsion. Besides, the average C-C length of C$_{24}$ fullerene does not vary much after confining H$_4$O inside.

**Table 1** Main geometrical characters of the optimized structure, the lowest vibrational frequencies (*f*), and HOMO-LUMO gaps of H$_4$O@C$_{24}$, H$_2$O, and C$_{24}$ at PBE0-D3/def2tzvp. (Unit: Angstrom for length; Degree for angle, cm$^{-1}$ for frequency, eV for energy).

| | $R_{O-H_1}$ | $R_{O-H_2}$ | $R_{O-H_3}$ | $R_{O-H_4}$ | | |
|---|---|---|---|---|---|---|
| **H$_4$O@C$_{24}$** | 0.996 | 0.997 | 1.021 | 1.021 | | |
| | ∠$H_1$-O-$H_2$ | ∠$H_1$-O-$H_3$ | ∠$H_1$-O-$H_4$ | ∠$H_2$-O-$H_3$ | ∠$H_2$-O-$H_4$ | ∠$H_3$-O-$H_4$ |
| | 110.74 | 112.35 | 105.94 | 106.45 | 112.17 | 109.27 |
| **H$_2$O**[a] | $R_{O-H}$ | | ∠H-O-H | | *f* | 180.97 |
| | 0.962 | | 104.18 | | | |
| **HOMO-LUMO gap of H$_4$O@C$_{24}$** | | | | 2.05 | | |

[a]calculated at CCSD(T)/aug-cc-pVTZ[26]

To show the bonding nature of the O-H bond of H$_4$O@C$_{24}$, localized molecular orbitals (LMOs) obtained through two localization methods including Foster-Boys and Pipek-Mezey localization method. The molecular orbital topology via Foster-Boys localization method of Fig. 3 as well as the orbital composition in Table S2 show that there exist four LMOs populating along the four O-H bonds of H$_4$O@C$_{24}$, which indicates there exists four electron pairs responsible for these four LMOs. H$_4$O@C$_{24}$ owns 156 electrons in total, which include 144 electrons from 24 C atoms, 4 electrons from 4 H atoms, and 8 electrons from O atom. Principally, as listed in Table S3, there should exist 48 LMOs to populate the electrons for C-C bonds of C$_{24}$ because it owns 24 single bonds and 12 double bonds; it needs 6 LMOs to populate the 12 electrons brought by the H$_4$O species. However, orbital analysis (Figure S2) shows that there exists 49 LMOs responsible for the C-C bonds and only exists 5 LMOs populating on the inner encaged H$_4$O, which implies that 2 electrons transfer from encapsulated H$_4$O to the outer C$_{24}$. It is similar as the condition of buckyball difluoride F$_2$@C$_{60}$ compound[8] and M@C$_{50}$ Series (M = Sc, Y, La, Ti, Zr, and Hf)[27]. Besides the evidence of number of LMOs, variations for the bond length for the related C-C bond between H$_4$O@C$_{24}$ from and C$_{24}$ also confirm the electron transfer as in Figure S3. In fullerene C$_{24}$, six single bonds and six double C-C bonds in turn connect the middle 12 C atoms, which can be seen from the corresponding bond lengths (1.356 and 1.452 Å in turn). However, for the C$_{24}$ of H$_4$O@C$_{24}$, results on the bond length how there are 7 C-C double bonds and 5 C-C single bonds among the middle 12 C atoms. As in Figure S3, the two pairs of successive C=C bonds with green background share similar bond length (1.441, 1.422, 1.440, and 1.423 Å) with those (bond length: ~1.460 Å) of the top and bottom C$_6$ cycle. In addition, natural population analysis shows that the charge of H$_4$O of H$_4$O@C$_{24}$ part is 1.16 *e* and whose value for C$_{24}$ of H$_4$O@C$_{24}$ -1.16 *e*, which also supports the electron transfer from inner encaged H$_4$O to outer C$_{24}$. Upon the electron transfer, only 10 electrons populate on H$_4$O, which includes the 1$s^2$ orbitals of the O atoms and the four orbitals for the O-H bonds.

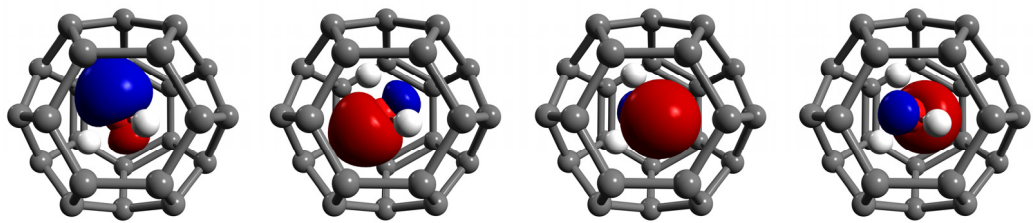

**Fig. 3** Four localized molecular orbitals obtained via Foster-Boys localization method for the four O-H bonds of H$_4$O@C$_{24}$. The blue lump corresponds to regions of space where the phase of the wave function is positive, and the red lump corresponds to regions of space where the phase of the wave function is negative.

The fullerenes with larger size were also examined. Figure S4 shows that C$_{26}$, C$_{28}$, and C$_{30}$ can also be used as nanoreactor for the generation of H$_4$O from H$_2$O and H$_2$ within 0.1 *ps*. The structure of H$_4$O@C$_{26}$, H$_4$O@C$_{28}$, and H$_4$O@C$_{30}$ optimized via PBE0-D3/def2tzvp are given in Fig. 4. Table 2 suggests that the bond length of the corresponding O-H bonds of H$_4$O@C$_{26}$, H$_4$O@C$_{28}$, and H$_4$O@C$_{30}$ resemble to those of H$_4$O@C$_{24}$. Localized molecular orbitals responsible for the four O-H bonds presented in Fig. 4 suggest that the O-H bonds own covalent characteristics. Table 2 shows that the lowest vibrational frequencies are 246.54 cm$^{-1}$, 212.82 cm$^{-1}$, and 322.24 cm$^{-1}$ for H$_4$O@C$_{26}$, H$_4$O@C$_{28}$, and H$_4$O@C$_{30}$, which means they own all positive frequencies indicating their stability. To show the chemical stability of H$_4$O@C$_{26}$, H$_4$O@C$_{28}$, and H$_4$O@C$_{30}$, their HOMO-LUMO gaps are given in Table 2. Results show that H$_4$O@C$_{26}$, H$_4$O@C$_{28}$, and H$_4$O@C$_{30}$ own larger HOMO-LUMO gaps, which indicates their high chemical stabilities. Similar to the case of H$_4$O@C$_{24}$, electron transfer also proceed in H$_4$O@C$_{26}$, H$_4$O@C$_{28}$, and H$_4$O@C$_{30}$ because there should exist 52, 56, and 60 LMOs populating on the outer fullerene cage. However, Table S4 and Figure S2 show that both the Foster-Boys localization method and Pipek-Mezey localization method suggest there actually exist 53, 57, and 61 LMOs populating on the outer fullerene, which means two electrons transfer from the inner $H_4O$ to the outer fullerene. Individual ab initio molecular dynamics at 298 K were also conducted for $H_4O@C_{26}$, $H_4O@C_{28}$, and $H_4O@C_{30}$ to further check their stability. Figure S5 shows that all the O-H bonds keep steady with 10 *ps* and the O atom are roughly close to the center of $C_{26}$, $C_{28}$, and $C_{30}$.

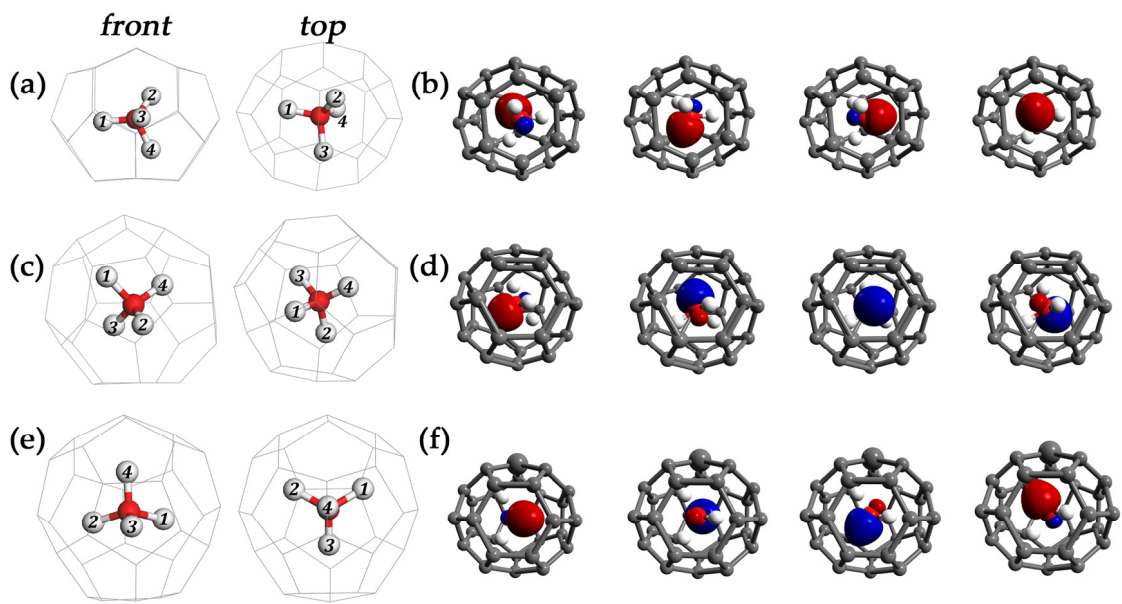

**Fig. 4** Optimized structure of (a) $H_4O@C_{26}$, (c) $H_4O@C_{28}$, and (e) $H_4O@C_{30}$ obtained via the PBE0-D3/def2tzvp level of theory. Localized molecular orbitals responsible for the four O-H bonds of (b) $H_4O@C_{26}$, (d) $H_4O@C_{28}$, and (f) $H_4O@C_{30}$ obtained via Foster-Boys localization method. The blue lump corresponds to regions of space where the phase of the wave function is positive, and the red lump corresponds to regions of space where the phase of the wave function is negative. Red: O atom; White: H atom; Grey: C atom.

**Table 2** Main geometrical characters of the optimized structure, HOMO-LUMO gaps, and the lowest vibrational frequencies (*f*) of $H_4O@C_{26}$, $H_4O@C_{28}$, and $H_4O@C_{30}$ calculated at the PBE0-D3/def2tzvp level of theory. (Unit: Angstrom for length; Degree for angle, $cm^{-1}$ for frequency, eV for energy).

|  | H$_4$O@C$_{26}$ | H$_4$O@C$_{28}$ | H$_4$O@C$_{30}$ |
|---|---|---|---|
| $R_{O-H_1}$ | 0.99 | 1.01 | 0.99 |
| $R_{O-H_2}$ | 1.02 | 1.04 | 1.02 |
| $R_{O-H_3}$ | 1.03 | 1.04 | 1.10 |
| $R_{O-H_4}$ | 1.06 | 1.09 | 1.11 |
| $f$ | 246.54 | 212.82 | 322.24 |
| **H-L gap** | 1.50 | 1.36 | 2.97 |

To further understand the interaction between H$_4$O and C$_{24}$ fullerene, energy decomposition analysis of H$_4$O@C$_{24}$ was performed at PBE0-D3/def2-TZVP level of theory. As shown in Fig. 5, the short intermolecular separation between H$_4$O and C$_{24}$ leads to high repulsion of them, which can be reflected through the repulsion ($E_{rep}$) term (112.72 eV). This term in turn contributes considerably to the formation of H$_4$O@C$_{24}$ through imposing constraint on the formation of the four O-H bonds. In the study of M@C$_{50}$ Series (M = Sc, Y, La, Ti, Zr, and Hf), the authors reported the same order of high repulsion energy of 37.30-48.44 eV for these six elements.[27] These values are smaller than that of this work because smallest radius of the adopted C$_{50}$ is around 3.10 Å[27] and the radii of these six elements range from 2.63 Å to 2.74 Å,[28] which means the separation of the carbon cages and encaged species is larger for the M@C$_{50}$ Series (M = Sc, Y, La, Ti, Zr, and Hf). The electrostatic term $E_{ele}$ (-34.45 eV), the exchange term $E_{ex}$ (-29.20 eV), the polarization term $E_{pol}$ (-40.69 eV), and the dispersion term $E_{disp}$ (-5.11 eV) together help to stabilize H$_4$O@C$_{24}$ complex through contributing attractive energy up to -109.45 eV, which leads to the total interaction of 3.27 eV. In comparison, CMO-EDA calculations show that the total interaction energies between H$_4$O and outer fullerene cage gradually decrease with the size increase of fullerenes, i.e. 0.98 eV for H$_4$O@C$_{26}$, -1.12 eV for H$_4$O@C$_{28}$, and -1.84 eV for H$_4$O@C$_{30}$. Notably, H$_4$O@C$_{28}$ and H$_4$O@C$_{30}$ present negative interaction energy. Upon the size increment of the fullerene, Fig. 5 indicates that the repulsion energy decreases considerably from 112.72 eV to 100.2, 89.59, and 80.49 eV due to the enlargement of the C-H distance. The population range between the outer C atom and the H atom increase from $C_{24}$ to $C_{26}$, $C_{28}$, and $C_{30}$ as in Figure S6, which leads to the decrease of the repulsive energy. In the meaning time, the attractive energies including dispersion energy, exchange energy, electrostatic energy, and polarization energy show that the strength for the attractive energies gradually decrease from $C_{24}$ to $C_{26}$, $C_{28}$, and $C_{30}$ and their respective percentages do not alter much for different fullerenes. This is consistent to the enlargement of the outer fullerene. Figure S6 also suggest that the most plausible C-H distance is still smaller than around 1.85 Å in $H_4O@C_{30}$, which is consistent to the reported radius of the C atom (1.90 Å)[28]. Actually, other larger fullerene like $C_{32}$, $C_{34}$, $C_{36}$ was also tested, which turns out not able to provide suitable confinement space for the generation of plausible structure of $H_4O$ as in Table S4.

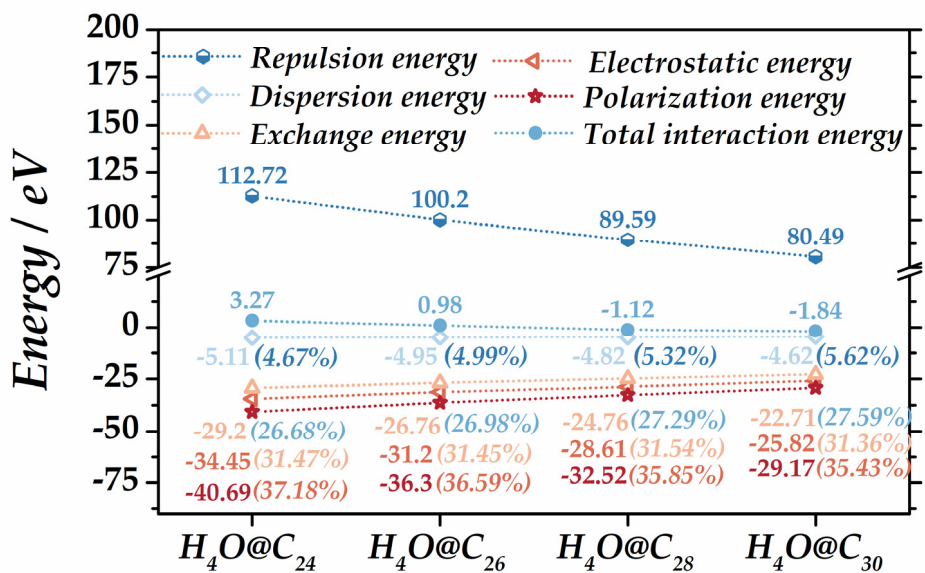

**Fig. 5** Pair interaction by CMO-EDA in $H_4O@C_{24}$, $H_4O@C_{26}$, $H_4O@C_{28}$, and $H_4O@C_{30}$. Units: eV

Analysis based on PBE0-D3/def2tzvp level of theory shows that that upon putting $H_4O$ inside of $C_{24}$, the corresponding C-C bonds were averagely lengthened from 1.446 to 1.492 Å with the maximum and minimum alternations of 0.094 and 0.002 Å as given in Table 3 and Figure S7, which is similar to the conclusion of previous work[29]. For $C_{26}$, $C_{28}$, and $C_{30}$, the averaged C-C length increase around 0.037, 0.025, and 0.021 Å at this level of theory. Following the method of dividing the fullerene into several tetrahedrons[30], the increment of the volume of the fullerene brought by the insertion of $H_4O$ were also analyzed. As in Table 3, the volumes of these four fullerenes increase 3.56, 3.20, 2.21, and 2.34 Å$^3$. The change of averaged C-C bond length and volumes from $C_{24}$ to $C_{26}$, $C_{28}$, and $C_{30}$ show a narrowing trend, which implies the deformation of the fullerenes generally decrease.

**Table 3** Averaged C-C bond length(Units: Å) and volumes (in parentheses; Units: Å$^3$) of $C_{24}$, $C_{26}$, $C_{28}$, and $C_{30}$ with or without the inner $H_4O$.

|  | Without $H_4O$ | With $H_4O$ | Δ |  | Without $H_4O$ | With $H_4O$ | Δ |
|---|---|---|---|---|---|---|---|
| $C_{24}$ | 1.492 | 1.446 | 0.046 | $C_{26}$ | 1.454 | 1.433 | 0.037 |
|  | (31.96) | (35.52) | (3.56) |  | (33.84) | (37.04) | (3.20) |
| $C_{28}$ | 1.463 | 1.438 | 0.025 | $C_{30}$ | 1.477 | 1.440 | 0.037 |
|  | (43.36) | (45.57) | (2.21) |  | (48.71) | (51.05) | (2.34) |

It can be anticipated that the formation of $H_4O@C_n$ via $H_2 + H_2O$ with fullerenes needs to absorb energy. To evaluate the energetic preference of the reaction of $H_2 + H_2O + C_n = H_4O@C_n$ (n = 24, 26, 28, and 30), their Gibbs energy changes were calculated as in Table 4. Results shows that with the increment of the size of fullerenes from $C_{24}$ to $C_{30}$, the energy needed for this type of reaction decrease from 14.03 to 7.29 eV. The energy change as well as the size of fullerenes suggest that the synthesis of $H_4O@C_n$ (n = 24, 26, 28, and 30) may not feasible for typical chemical synthesis protocol, but is possible with beam implantation method[31] or other advanced experimental technique for the penetration of $H_2O$ followed by hydrogen tunneling. Previous studies have found that Ne with similar radius (1.56 Å) with O atom (1.71 Å) can penetrate into/out $C_{60}$ [31, 32], which implies $H_2O$ could have a chance of penetrating into the interior of fullerene on colliding with the cage molecules. Although calculation shows that $H_4O^{2+}$ can exist in free space with no imaginary frequency, this work focuses on the unexpected synthesis in fullerene, i.e. the claimed nano-reactor, for species (e.g. $H_2$ and $H_2O$) that cannot react under ambient conditions.

**Table 4** Gibbs energy change for reaction of $H_2 + H_2O + C_n = H_4O@C_n$ (n = 24, 26, 28, and 30)

| Reaction | Gibbs energy change /eV |
|---|---|
| $H_2 + H_2O + C_{24} = H_4O@C_{24}$ | 14.03 |
| $H_2 + H_2O + C_{26} = H_4O@C_{26}$ | 11.18 |
| $H_2 + H_2O + C_{28} = H_4O@C_{28}$ | 8.19 |
| $H_2 + H_2O + C_{30} = H_4O@C_{30}$ | 7.29 |

4. CONCLUSION

In summary, we theoretically proved the hyper-hydrogenated water species $H_4O$ can be synthesized from $H_2$ and $H_2O$ and stably maintained in the fullerenes at ambient condition. This work proposes a potential synthesis protocol for generating nonexistent molecules in future.

ASSOCIATED CONTENT

**Supporting Information**. The following files are available free of charge. The number of electrons and orbitals of $H_4O@C_{24}$, $H_4O@C_{26}$, $H_4O@C_{28}$, $H_4O@C_{30}$; Comparison of C-C length for single $C_{24}$ and $C_{24}$ from $H_4O@C_{24}$; Key bond lengths during AIMD; Coordinates and frequencies obtained via PBE0-D3/def2tzvp for $H_4O@C_{24}$, $H_4O@C_{26}$, $H_4O@C_{28}$, $H_4O@C_{30}$; Figure S1-S7; Table S1-S4. (PDF)

# AUTHOR INFORMATION

**Notes**

The authors declare no competing financial interests.

# ACKNOWLEDGMENT

This work was supported by the National Natural Science Foundation of China (21773287 and 12204213). The computations were performed at the National Supercomputing Center in Guangzhou (NSCC-GZ) and Shanghai.